\documentclass[preprint2]{emulateapj}

\usepackage{url}

\RequirePackage{color}
\usepackage{amssymb}
\usepackage{color}
\usepackage{comment}
\usepackage{graphicx}

\def\cm{\,{\rm cm}}

\def\ergscm2 {erg\,s$^{-1}$cm$^{-2}$}

\def\cm2 {cm$^{-2}$}

\def\aap {A\&A}
\def\apj {ApJ}
\def\apjl {ApJ Lett.}
\def\apjs {Astrophys. J. Suppl. Ser.}
\def\epl{Euro. Phys. Lett.}
\def\ssr{Space Sci. Rev.}

\shorttitle{Nuclear Fusion in the Deuterated cores of inflated hot Jupiters}
\shortauthors{Ouyed\&Jaikumar}

\begin{document}


\title{Nuclear Fusion in the Deuterated cores of inflated hot Jupiters}

\author{Rachid Ouyed$^{1}$\thanks{Email:rouyed@ucalgary.ca} and Prashanth Jaikumar$^{2}$}

\affil{$^1$Department of Physics \& Astronomy, University of Calgary, 2500 University Drive NW, Calgary, AB T2N 1N4, Canada}

\affil{$^2$Department of Physics and Astronomy, California State University  Long Beach,  1250 Bellflower Blvd., Long Beach CA 90840}

\begin{abstract}
Ouyed et al. (1998) proposed Deuterium (DD) fusion at the core-mantle interface of giant planets as a mechanism to explain their observed heat excess.
 But rather high interior temperatures ($\sim 10^5$ K) and a stratified D layer are needed, making such a scenario unlikely. In this paper, we re-examine DD fusion, with the addition of screening effects pertinent to a deuterated core containing ice and some heavy elements. This alleviates the extreme temperature constraint and removes the requirement of a stratified D layer. As an application, we propose that,  if their core temperatures are a few times $10^4$ K and core composition is chemically inhomogeneous, the observed inflated size of some giant exoplanets (``hot Jupiters") may be linked to screened DD fusion occurring deep in the interior. Application of an analytic evolution model suggests that the amount of inflation from this effect can be important if there is sufficient rock-ice in the core, making DD fusion an effective extra internal energy source for radius inflation. The mechanism of screened DD fusion, operating in the above temperature range, is generally consistent with the trend in radius anomaly with planetary equilibrium temperature $T_{\rm eq}$, and also depends on planetary mass. Although we do not consider the effect of incident stellar flux, we expect that a minimum level of irradiation is necessary to trigger core erosion and subsequent DD fusion inside the planet. Since DD fusion is quite sensitive to the screening potential inferred from laboratory experiments, observations of inflated hot Jupiters may help constrain screening effects in the cores of giant planets. \end{abstract}
~\\~\\

\section{Introduction}

The natural assumption in evolutionary models of Jupiter-size exoplanets that are as old as their stellar hosts (several Gyr) is that they should have evolved to their degeneracy dominated configuration, their radius converging to a Jupiter-like radius $R_{\rm J}$ (e.g. Zapolsky \& Salpeter 1969; Stevenson 1982; Hubbard 1984).   But the observed size of such exoplanets belies this modeling assumption, even when strong irradiation from the parent star is taken into account\footnote{For Jupiter-mass planets parked at 0.02 AU, radii of $\sim1.2 R_{\rm J}$ are expected at Gyr ages (e.g Burrows et al. 2007). Nevertheless,  many planets have radii in excess of $1.2 R_{\rm J}$, and most receive irradiation below that expected at 0.02 AU (e.g. Miller et al. 2009).}. This ``radius anomaly" of hot Jupiters (Guillot et al. 2006), combined with the variation in radii of most known exoplanets, may be considered a challenge to standard evolutionary models of these objects.

A few explanations put forward are : (i) capture of a small percentage of the stellar radiation, or wind, that is deposited deep inside the hot exoplanet, providing an  ``internal heat" source to slow down the cooling sufficiently to inflate the planet; (e.g., Baraffe et al. 2010); (ii) tidal heating due to unseen companions pumping up the eccentricity (Bodenheimer et al. 2001; see also Jackson et al. 2008; 
 Hansen 2008; Miller et al. 2009; Ibgui \& Burrows 2009); (iii) kinetic heating due to the breaking of atmospheric waves (Guillot \& Showman 2002) or magneto-viscous heating from ionized winds in the magnetized upper atmosphere (Batygin \& Stevenson 2010); (iv) enhanced atmospheric opacity (Burrows et al. 2007); (v) layered convection induced by compositional inhomogeneity which slows the contraction due to inefficient heat transport in the interior (Chabrier \& Baraffe 2007).  A review of these ideas can be found in Fortney \& Nettelman (2010). Huang \& Cumming (2012) found it challenging to inflate massive planets or those with strong irradiation using Ohmic heating alone.  Burial of heat by turbulence (Youdin \& Mitchell 2010), and strong winds (Chabrier et al. 2011) are other mechanisms invoked to diffuse stellar irradiation into the planet interior, though recently, Kurokawa \& Inutsuka (2015) argued that layered convection with a monotonic chemical composition gradient may be insufficient to explain the amount of radius inflation. It has also been proposed that very short period, low-mass, binaries could be the progenitors of stellar mergers which may lead to the observed population of very hot Jupiters (Martin et al. 2011). In this model, inflation is  an indication of youth.

As a result, while the radius anomaly of hot jupiters probably originates from additional sources of energy dissipation, the details of the mechanism for radius inflation are a topic of current debate. 
 
 In light of this, we are motivated to explore in more detail the Deuterium fusion model  of Ouyed et al. (1998; hereafter OFCS98), which we present here as a feasible alternative scenario that may be an important factor in (or even at the root of) the radius anomaly. As this is a truly deep internal nuclear source of energy generation, not one that traps the external irradiation through atmospheric phenomena, our idea is novel within the range of present explanations for inflated hot jupiters.
 
 In  OFCS98, we pursued the possibility that the excess heat of Jupiter might be due to DD fusion in a sedimented layer of pure D only a few km in thickness and concentration $\sim$ 2-4 g cm$^{-3}$. Extrapolating the DD fusion rates from laboratory energies down to the scale of planetary cores, we found that a temperature as high as $T \sim 10^5$ K is needed to explain the excess heat, which is an order of magnitude larger than normally expected core temperatures. Furthermore, the sedimentation and stratification  of the D layer required an unspecified mechanism. The stability of the layer to convection also posed a challenge to this picture. Here, we present a modification to the original idea in OFCS98 that alleviates these issues and apply it to Jovian exoplanets.  By shifting the source of D from the inner layers of the envelope to the deuterated core and including an effective screening potential, we argue that energy release from DD fusion at typical core temperatures of  $\sim 10^4$ K is still sufficient to slow down the cooling of the exoplanet. With resulting luminosities as large as $P_{\rm DD}\approx 10^{26}$ erg s$^{-1}$, screened DD fusion in the deuterated core can provide a viable persistent, energy source to inflate close-in, hot Jupiters. 

 Let us recall that a viable mechanism to explain the excess heat and radius anomaly should be consistent with the following key facts:   

 \indent    {\bf (i)} The energy release must be persistent and must be operating on timescales much larger 
    than the Kelvin-Helmholtz times ($\tau_{\rm KH} \sim 10^8$ yrs; the time it takes a  Jupiter-like planet to settle back to its degeneracy dominated configuration); \\
 \indent      {\bf (ii)} The data shows that the radius anomaly $\Delta R$=$R_{\rm obs}$-$R_{\rm pred}$ (the difference between the
 observed and predicted radius) scales with the
          planetary equilibrium temperature $T_{\rm eq}$, with exponents that vary with the mass range. While Laughlin et al. (2011) initially found a global best-fit dependence $\Delta R \propto T_{\rm eq}^{\alpha} \,{\rm with}\, \alpha$=$1.4\pm 0.6$,  subsequent work (Enoch et al. 2012; Weiss et al. (2013)) has found that taking planet mass, incident flux (or orbital separation) into account provides more accurate fits for subsets of the data. Weiss et al. (2013) have argued that incident flux is a better indicator of radius than planetary mass for jupiter-mass planets, and their fits support $\Delta R \propto T_{\rm eq}^{\alpha} \,{\rm with}\, \alpha$=$0.5$. Their data also show that for smaller mass planets, the radius does depend strongly on the mass. A proposed mechanism should include and investigate this mass-radius relationship.\\
          Guillot et al (2006) had proposed that the radius-inflation mechanism is anti-correlated with the parent star metallicity, i.e,
        the radius anomaly grows smaller with increasing metallicity. However, this is not a strong constraint since
        if the planets around metal-rich stars are themselves metal-rich, they would be naturally be denser and more compact (Johnson et al. (2010); Enoch et al. (2012)).
        The issue is still debated as the correlation between parent star metallicity and planet metallicity is weaker than originally thought (Thorngren et al. (2015)). We therefore do not take the metallicity trends as a significant constraint on the inflation mechanism.

The above two points can be obtained in our model if we restrict the parameter space of the screening potential based on the astrophysics.
This paper is organized as follows: In \S 2 we describe screened DD burning under conditions
   prevailing in the interior of  hot, close-in, Jupiters and estimate the screened reaction rate at relevant temperatures. In \S 3, we apply the results to the core of hot jupiters, estimating 
    the fusion power from this reaction, the amount of radius inflation from a simple evolution model, timescale of the effect, the trend with planetary equilibrium temperature $T_{\rm eq}$ and the trend with planetary mass.
       Concluding remarks are presented in \S 4.

\section{DD fusion in Jovian exoplanets}

  \subsection{DD fusion in a D layer}
  \label{sec:PDD}
  
  In OFCS98, assuming a concentrated layer 
 of D near or above the core in giant planets,  we showed that DD burning can be triggered and sustained if
  sufficiently high temperatures ($T> \sim 10^5$ K) are maintained once the D layer forms\footnote{The Deuterium
  could be supplied by planetesimal vaporization (Ouyed 2004) or, as we argue in this paper, by core erosion (e.g.
  Guillot et al. 2004).}. At $T< 10^5$ K (i.e. $T<  \sim 6$ eV), the DD cross-section decreases sharply  (Fig. 1 in OFCS98). 
   These are extreme conditions since (i) a $10^5$ K temperature may barely be achieved, if at all, only early
   in the planet's history; (ii) the formation, stability  and preservation of the D layer is compromised by the strong convective mixing in the interior.  
  In OFCS98, we therefore envisaged the D layer to be situated in a non-convective portion
 of the planet's interior.   At a temperature $T$,  the reactivity $\langle \sigma v \rangle$ used in OFCS98 is the Bosch-Hale fit (Bosch \& Hale 1992) based on the functional form suggested by Peres (1979): $\langle \sigma v \rangle = C_1 \theta \sqrt{\frac{\zeta}{\mu c^2 T^3}}  e^{-3\zeta}$ cm$^3$ s$^{-1}$ where 
   $\theta$ and $\zeta$ are given by eqs. (6) and (7) respectively in OFCS 98\footnote{Eq.(6)  in OFCS98 should read $\theta = \frac{T}{1- \frac{T(C_2+T(C_4+TC_6))}{1+ T (C_3+ T(C_5+TC_7))}}$ with $T$ in keV and fitting parameters $C_{\rm i}$ from  Bosch \& Hale (1992).} while $\mu$ is the reduced mass of the reactants and $c$ the speed of light. The exponential fall of the reactivity, some 10 orders of magnitude for a factor
   of 2 in temperature, is moderated by standard electron and ion screening effects so that the effective reaction rate $\sigma_{\rm eff} = f_{\rm e} f_{\rm i} \langle \sigma v \rangle$. Here, $f_{\rm e}(T, \rho)$ is the strong screening limit of the degenerate quantum electron gas  while $f_{\rm i}  (T)$ accounts for screening by the classical ionic liquid formed by D nuclei (eqs. (8) \& (9) respectively in OFCS98).
   The power generated from the D layer was found to be of the order $10^{24}$ erg s$^{-1}$ for temperatures of $\sim 10^5$ K at densities
 prevailing in the core-mantle region inside planets such as Jupiter. 
   Once DD burning starts, the reactivity and thus the temperature in the D layer increases.
     Two possible outcomes are: (i) a runaway burning because of the degenerate core conditions. 
     In this case, most of the D is consumed
       on very short timescales (compared to the planet's lifetime) leading to a ``sudden" release of energy; (ii) a steady-state burning at  an equilibrium 
         temperature ($T_{\rm c}\sim 10^5$ K) below the convection threshold. This second scenario may be a viable explanation of the heat excess, but the high temperature and survival of the D layer once burning starts are unfavorable and adhoc conditions (Ouyed 2004).

    \subsection{DD fusion in the deuterated core}

To find a more plausible alternative to these extreme conditions, we take a closer look at DD screening in a deuterated core environment.
Here, DD burning is occurring in the core which is  bombarded by the D ions freed from the icy D-bearing 
core material by erosion (e.g. Guillot et al. 2004).   The freed D which is lifted upward by convection 
eventually settles back down onto the core triggering screened fusion. Thus we rely on the presence of icy material
in the core to supply the free D (the projectiles or beam) which fuse with the D in the core (the target). 
Other sources and supply mechanisms of free D (which we do not consider here) 
originate in the overlaying H-rich envelope and deposition of D 
 deep inside the planet's interior by planetesimal impact and vaporization (Ouyed 2004).

 Let us represent the screened cross-section by $\sigma_{\rm scr}$. Several laboratory-based studies have shown that the DD fusion cross-section is enhanced in a solid deuterated target as compared to a gas target (Czerski et al. 2001, Raiola et al. 2002, Kasagi et al. 2002, Kasagi  2004). This is attributed to correlations between the conduction electrons in a metal lattice and enhanced mobility of deuterons. The environment of the target nucleus, in particular, the mobility of D ions in the metal lattice, plays an important  role in fusion well below the Coulomb barrier, and is the principal reason for the large enhancement. The enhancement factor at energy $E$ is (Raiola et al. 2002, Kasagi  et al. 2004) 
\begin{equation}
\label{enhance}
f_{\rm scr}=\frac{\sigma_{\rm scr}}{\sigma_{\rm bare}}=\frac{E}{E+U}\frac{{\rm Exp}(bE^{-1/2}-1)}{{\rm Exp}(b(E+U)^{-1/2}-1)}\ ,
\end{equation}
 for $S(E)\approx S(E+U)$ where $b=\pi\alpha_e\sqrt{2\mu c^2}$ with $\alpha_e$ the fine structure constant, $U$ is the screening potential for the metal target, and $S$ the astrophysical S-factor. 
 The value of the screening potential $U$ has a large impact on our calculations, since the DD fusion cross-section is exponentially sensitive to it at low temperatures. Laboratory values for $U$ vary widely from one target to the next, even within the same experimental setup (eg., from $\leq 30$ eV for Al target to $\approx 450\pm 50$ eV for Fe target; Raiola et al. 2002) and have not achieved consensus between different groups (Kasagi 2004 reports  $U\approx 200\pm 20$  eV for Fe), so that we cannot choose a value with confidence {\it a priori}. In addition, the absence of detailed knowledge of the core composition affects the choice of $U$. While laboratory experiments do extremely well to achieve a few hundred eV precision for nuclear physics (MeV scales), this is insufficient to constrain our model. We will therefore allow the astrophysics to further constrain the range of $U$ by requiring that the resulting power from DD fusion $P_{\rm DD}$ allows 
  some amount of inflation while lasting for Giga-years. We emphasize that the sensitivity to $U$ is an opportunity to use exoplanet observations to constrain the screening effects in a unique regime of temperature and density. Adopting a typical value of $U$ of 200($\pm$20\,eV) below makes our model consistent with observational numbers and trends (see sec.\S3), and this value corresponds to the screening parameter in a deuterated Fe target, as per Kasagi et al. (2004). 
 Since D is expected to be found in ice pockets, this could be interpreted to mean that Fe is the main component of the rocky material that diffuses from deeper layers and gets mixed with ice.\\
 
Taking screening into account, we can write the corresponding thermally averaged cross-section of DD fusion as
\begin{equation}
\langle \sigma_{\rm scr}v\rangle=\int_0^{\infty}\sigma_{\rm scr}(v)\,vf(v)dv \ ,
\label{thermal}
\end{equation}
where $v=\sqrt{2E/\mu}$ and $f(v)$ is the Maxwellian distribution of speeds. The integral is usually carried out in the energy domain by approximating the integrand as a narrow Gaussian around the Gamow peak. The S-factor is roughly energy independent, and the thermally averaged rate can then be extrapolated to temperatures of astrophysical interest based on fits to multi-channel cross-sections from R-matrix theory (Bosch \& Hale 1992). In the case of extrasolar Jupiters, core temperatures are in the range of a few eVs, much lower than in stellar cores where temperatures are of order keV.  A naive extrapolation of the Bosch-Hale fit to $k_{\rm B} T\sim$ eV shows that $\theta\lesssim T$ (see footnote 3), so the rate falls off at least as rapidly as $T^{-2/3}{\rm exp}(-3E_0/k_BT)$ where $E_0$=$(bk_BT/2)^{2/3}$; here $k_{\rm B}$ is the Boltzmann constant. However, now the screening enhancement boosts the reaction rate by several orders of magnitude at low temperature. This effect is immediately apparent when we approximate $E/(E+U)\approx E_{\rm pk}/U$ in the pre-factor of Eq.~(\ref{enhance}) where $E_{\rm pk}$ is obtained from
\begin{equation}
\frac{d}{dE}\left(\frac{b}{\sqrt{E+U}}+\frac{E}{k_BT}\right)\biggl|_{E=E_{\rm pk}}=0\ .
\label{oldpeak}
\end{equation}

\begin{figure}[t!]
\centering
\includegraphics[scale=0.65]{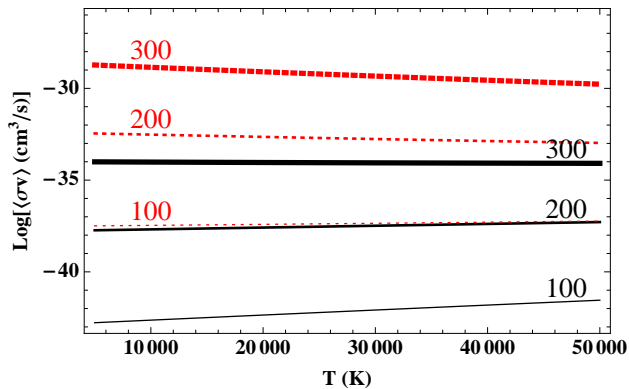}
\caption{{Reactivity $\langle\sigma_{\rm scr} v\rangle$ (solid lines) from Eq.~(\ref{thermal}) for different values of screening potential $U$ from 100-300 eV, and the corresponding effective cross-section $\sigma_{\rm eff} = f_I(T)\langle\sigma_{\rm scr}\,v\rangle$ (dotted line) for $5\times 10^3\ {\rm K} \le T \le 5\times 10^4\ {\rm K}$, representative of the core of giant planets. The value of $U$=200$\pm20$ eV for Eq.~(\ref{enhance}) corresponds to an Fe target (Kasagi 2004) and appears to fit best with our model.}~\\}
\label{fig-sigma}
\end{figure}

We find that $E_{\rm pk}=E_0-U$, which adds a new multiplicative factor ${\rm exp}(U/T)$ in the Gaussian approximation to the rate - this explains the strong sensitivity of the cross-section to the value of $U$ apparent from Fig.~\ref{fig-sigma}. Since $E_{\rm pk}\geq 0$ implies $E_0>U$, our estimate of the enhancement factor is accurate for temperatures that are large enough to satisfy $k_BT > U^{3/2}(2/b)$ (typically 5 eV or more if we choose $U\approx 200$ eV). Taking $U\rightarrow 0$ yields the usual temperature dependence of the reactivity in the absence of screening: ${\rm exp}(-3E_0/k_BT)$. But core temperatures in extrasolar Jupiters can be somewhat lower than 5eV, so we make a more accurate calculation of $E_{\rm pk}$ valid for $E_0< U$ by setting
\begin{equation}
\frac{d}{dE}\left(\frac{b}{\sqrt{E+U}}+\frac{E}{k_BT}+{\rm ln}\left[\frac{E+U}{E}\right]\right)\biggl|_{E=E_{\rm pk}}=0\ .
\label{newpeak}
\end{equation}

The Gamow peak is now located quite close to $k_BT$, which tempers the exponential fall-off with temperature. As before, we perform a Gaussian approximation to the integrand to evaluate the integral. The reaction rate $\langle\sigma_{\rm scr} v\rangle$ now behaves as $\sim T^{1/2} {\rm exp}({-b/\sqrt{U}})\times {\rm exp}\biggl(\frac{E_0^{3/2}}{U^{3/2}}\biggr)$. While this is a monotonically increasing function of temperature, the effective cross-section $\sigma_{\rm eff}$=$f_I\langle\sigma_{\rm scr}\,v\rangle$ is monotonically decreasing because of the strong effect of ionic screening $f_I(T)$ at low temperatures. The factor $f_{\rm e}(\rho,T)$ need not be included in $\sigma_{\rm eff}$ since the screening due to conduction electrons is already present in $U$. The two approximate expressions for the rate $\sigma_{\rm eff}$, one rising and the other falling with $T$ on either side of $k_BT\approx U^{3/2}(2/b)$, are joined together by a cubic Hermite spline to obtain a smooth continuous curve for the rate around a few eVs. The two individual rates are naturally quite sensitive to $U$, therefore, we expect the fitted fusion reactivity to depend strongly on $U$ as well, and indeed this is what is seen in Fig.~\ref{fig-sigma}, which shows the reactivity $\langle\sigma_{\rm scr} v\rangle$ with the effective potential $U$ included, alongside the corresponding effective cross-section $\sigma_{\rm eff}=f_I(T)\langle\sigma_{\rm scr}\,v\rangle$ for $5\times 10^3\ {\rm K} \le T \le 5\times 10^4\ {\rm K}$, representative of the core of giant planets.

\section{Application to giant/hot planets/Jupiters}

 \subsection{Fusion Power}
 
\begin{figure}[t!]
\centering
\includegraphics[scale=0.65]{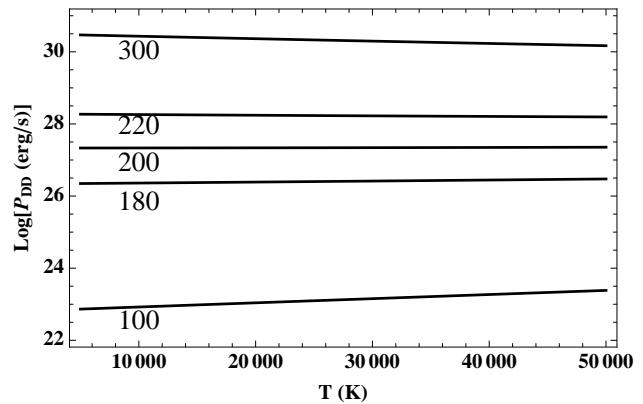}
\caption{{Fusion power Log($P_{\rm DD}$) from Eq.~(\ref{Pdd}) for different values of screening potential $U$ (in eV) for $5\times 10^3\ {\rm K} \le T \le 5\times 10^4\ {\rm K}$, representative of the core of giant planets. A value of $U\approx 180$-220 eV is appropriate for Fe-dominated rock-ice, which gives $P_{\rm DD}\approx 10^{26}$-$10^{28}$ erg s$^{-1}$, yielding sufficient inflation.}~\\}
\label{fig-pdd}
\end{figure}

Taking into account the two main channels of DD fusion, viz., d(d,n)$^3$He and d(d,p)t (see the DD cycle
and Fig.3 in OFCS (1998)), we compute the power release $P_{\rm DD}$ for a specific core density of
 D ($\rho_{\rm D}$=$n_{\rm D} m_{\rm D}$) and $U$ value\footnote{Although $P_{\rm DD}$ depends weakly on temperature $T$ through the reactivity $\sigma_{\rm eff}$, we do not write it explicitly since the dependence with $U$ is more important in the temperature range $10^4$-$10^5$K. In fact, $P_{\rm DD}$ for our median value of $U\approx$ 200 eV is almost constant in this temperature range (see Fig.~\ref{fig-pdd}).} as follows (see Eq.~(14) in OFCS98):
\begin{equation}
\label{eq:pdd1}
P_{\rm DD; i}(U)= \int_{0}^{R_{\rm c}} Q_i \frac{n_{\rm D}^2}{2} \sigma_{\rm eff} (U)4\pi r^2 dr\ ,
\end{equation} 
where $R_{\rm c}$ is the core radius and $m_{\rm D}$ is deuterium mass.
 The index $i=1,2$ stands for the two main channels of DD fusion with $Q_i$ the energy
released per D atom fused for each channel. 
 In terms of core density $\rho_{\rm c}$, the core density of D is  $\rho_{\rm D} = 2 (D/H)\times \rho_{\rm H}=  2 (D/H)\times  (2/18) \rho_{\rm ice}$, which gives
\begin{equation}
\rho_{\rm D}  \sim 2.2\times 10^{-6} \eta_{\rm D, -4} \eta_{\rm ice, -1} \rho_{\rm c}\ ,
\label{metals}
\end{equation}
where $\eta_{\rm ice, -1}$ is the amount in mass of ice in the core in units of $10^{-1}$ (i.e. $\rho_{\rm ice}$=$0.1\rho_{\rm c}$) and $\eta_{\rm D, -4}$ the D to H ratio (D/H)
in the core in units of  $10^{-4}$. The value of D/H ratio chosen is a few times higher than its protosolar value, since this isotopic ratio is higher in planetary ice than interstellar medium (Guillot (1999)). Within our model, one must address the likelihood that heavy elements from the rocky, icy core can be well-mixed with the less dense layers of H above it, so that the D liberated by core erosion finds itself in a metallic deuterated environment. Recent works have argued that heavy elements from the core can be mixed with fluid H/He layers at the core-mantle interface due to sedimentation or core erosion (Wilson \& Militzer 2012a; 2012b) leading to chemically inhomogeneous layers deep in the planetary interior. In these works, it was not clear that the solvated materials could be convected upwards efficiently through the lighter layers on top. However, Wilson (2015) has proposed that the mechanism of semi-convection can still lead to high diffusivity of the heavy rock-ice into the upper layers, based on modern density-functional calculations. Semi-convection separates layers of constant composition with thin diffusive layers that are well-mixed, leading to a slower rate of redistribution and a slower rate of cooling. This also implies that core erosion and mixing can persist for longer times. In such a scenario, our model becomes viable since DD fusion will occur in a deuterated metallic environment, supported by slow core erosion and slow cooling. Assuming a uniform spatial profile for the core density, we have

\begin{equation}
P_{\rm DD;i} (U)\sim  \frac{Q_i}{2} (2.2\times 10^{-6} )^2 \eta_{\rm D, -4}^2 \eta_{\rm ice, -1}^2 \frac{\rho_c}{m_{\rm D}} \frac{M_{c}}{m_{\rm D}}\sigma_{\rm eff}(U)\ ,
\label{Pdd}
\end{equation}
where $M_{\rm c}= \int_0^{\rm R_{\rm c}} \rho_{\rm c} 4\pi r^2 dr$ is the core mass.

The amount of energy released is $Q_{\rm DD}= Q_1+Q_2\simeq 2\times  10^{-5}$ ergs per D atom fused, with $P_{\rm DD}= P_{\rm DD, 1}+ P_{\rm DD,2}$.  The mass of the core is hereafter taken to be 10\% of the planet's mass ($M_c$=$0.1 M_{\rm P}$). For an evolved Jupiter size ($R_{\rm J}$) and Jupiter mass ($M_{\rm J}$) planet, with $\rho_{\rm c, J}\simeq 4.38$ g cm$^{-3}$ as core density (e.g. Hubbard 1984), we find $P_{\rm DD} (U)\sim  \left(3.8\times 10^{26}\frac{\rm erg}{\rm s}\right)\eta_{\rm D,-4}^2\eta_{\rm ice, -1}^2 \frac{\sigma_{\rm eff}(U)}{10^{-34}\ {\rm cm}^3{\rm /s}}$. 
 The effective fusion reactivity $\sigma_{\rm eff}(U)$  
   is scaled in units of its typical magnitude $10^{-34}$ cm$^3$ s$^{-1}$ for $U$=180 eV (typical of rock-ice dominated by Fe), and a burning temperature 
   in the $T_{\rm c}\sim 10^4$ K range.  We can get an idea of the sensitivity of $P_{\rm DD}(U)$ to the screening potential in Fig.~\ref{fig-pdd} which shows the result for $P_{\rm DD}$ as a function of core temperature.

\subsection{Amount of Radius Inflation}

To study the amount of inflation that results from DD fusion over Gyr timescales, it is useful to employ a one-dimensional analytic time-evolution model for the planetary radius $R_{\rm P}(t)$ (e.g.  Guillot 2005).  I.e. $dE/dt=P_{\rm DD}(t)-L_{\rm eff}(T_{\rm eff})$ where $E=-3GM_{\rm P}^2/7R_{\rm P}(t)$ is the (time-dependent) internal energy (we make use of the  virial theorem to link the planet's internal energy to gravitational
energy), and $L_{\rm eff}(T_{\rm eff}) = 4\pi R_{\rm P}(t)^2 \sigma_{\rm SB} T_{\rm eff}^4$ is the standard blackbody luminosity;
$\sigma_{\rm SB}$ being the Stefan-Boltzmann constant.  The time evolution of $P_{\rm DD}(t)$, which is the additional internal heat source, comes from the following factors: the core mass $M_c$ and density $\rho_{\rm c}$, and the effective cross section $\sigma_{\rm eff}$ which depends on the core temperature $T_c$.  We note from Fig.~\ref{fig-sigma} that although the core temperature may change by a factor of a few during the contraction, the cross-section is only weakly dependent on temperature for screening potential U typical of Fe-dominated rock-ice, so that its time-dependence can be ignored.

We express the core density in terms of the planetary mass and radius following the prescription given in Seager et al. (2007; see  Eq.(11)
and Table III in that paper) together with pressure balance between the core and the inner envelope.  This effectively 
 gives us the time-dependence of the core density in terms of $dR_{\rm P}/dt$. The core mass erodes at a rate proportional to the existing core mass (since erosion powers $P_{\rm DD}$) leading to $M_{\rm c}(t)=M_{\rm c}(0){\rm exp}(-0.1 t)$
 with $M_{\rm c}(0)=0.1 M_{\rm p}$. As time is measured in Gyr, the core mass effectively decreases by $\approx 10\%$ on a timescale of 1 Gyr and about 65\% in 10 Gyr.   We use the mean adiabat connecting the isothermal core at temperature $T_c$ to the effective T$_{\rm eff}$ which determines the cooling flux (ignoring the effect of stellar irradiation of the atmosphere). We find that $T_{\rm eff}$ scales as $M^{0.4}$ for a mean adiabatic index of $\Gamma$=0.3, as commonly used for Jupiter interiors (eg. Hubbard et al. 1968).  Here,  $T_{\rm eff}= (200\ {\rm K}) (M_{\rm p}/M_{\rm J})^{0.4}$.

\begin{figure}[t!]
\centering
\includegraphics[scale=0.65]{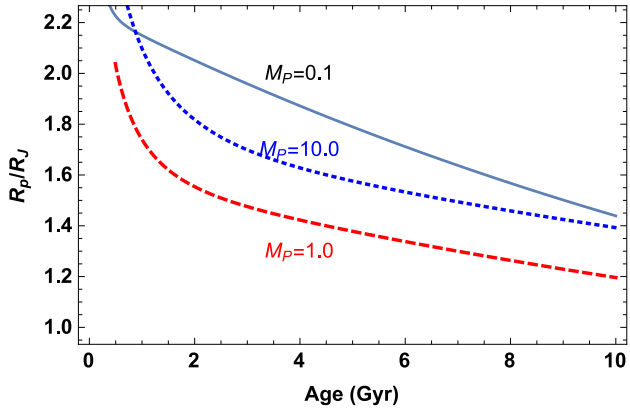}
\caption{{Time evolution of $R_{\rm P}(t)$ including DD fusion in the planetary core, assuming $\eta_{\rm ice}=0.1$ and no stellar irradiation. Masses $M_{\rm P}$ are in units of Jupiter mass. Sub-jupiter mass planets are inflated the most, due to smaller heat loss ($T_{\rm eff}$; see \S \ref{sec:masstrends}) while higher mass planets have larger internal heat from $P_{\rm DD}$ to balance cooling.}~\\}
\label{fig-inflation}
\end{figure}

The solution of the evolution equation for $R(t)$ is plotted in Fig.\ref{fig-inflation}, from which we infer that DD fusion can provide enough energy to prevent prolonged contraction, and ultimately lead to an equilibrium inflated radius that persists for longer than the Kelvin-Helmholtz timescale. We discuss the trend with planetary mass in \S \ref{sec:masstrends}.
 
\begin{figure}[t!]
\centering
\includegraphics[scale=0.65]{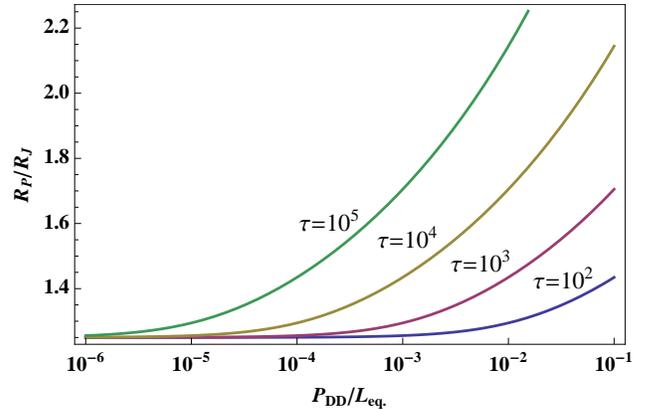}
\caption{{Scaling of planetary radius  (with respect to Jupiter radius) with DD fusion power (relative to equilibrium luminosity $L_{\rm eq}$)  for different values of the optical depth $\tau=\{10^2,10^3,10^4,10^5\}$ and corresponding pressure at the base of the radiative layer $P=\{1,10,100,1000\}$ bars}, using Eq.~(\ref{inflate}).~\\}
\label{fig-anomaly}
\end{figure}

Taking into account opacity effects in the atmosphere leads to even more inflation.
The semi-analytic treatment in Ginzburg \& Sari (2015) which expresses $\Delta R$ in terms of the deposited luminosity and optical depth for a deep internal heating source supports our numerical estimate from the radius evolution above. If we compare $P_{\rm DD}$ to $L_{\rm eq}$=$4\pi R_{\rm P}^2 \sigma_{\rm SB}T_{\rm eq.}^4\sim  10^{28}$-$10^{30}$ erg s$^{-1}$ (the planet's equilibrium luminosity and $T_{\rm eq.}$ the equilibrium temperature), we find that fusion power can be from a few percent (for $M_{\rm P}\sim M_{\rm J}$) up to 100\% (for $M_{\rm P}\sim 10 M_{\rm J}$) of $L_{\rm eq}$. This heat, being deposited near the core boundary, can be convected up rapidly through fluid motion and deposited at the base of the radiative layer, where thermal resistance is large. In this deposition region, opacity from negative Hydrogen ion H$^{-}$, $\kappa_{\rm H}$,  dominates the heat transfer at $T_{\rm eq}$
    of a few thousand Kelvins (e.g. Arras \& Bildsten 2006). Using Eq.~(31) in Ginzburg \& Sari (2015), we find
    \begin{equation}
    \Delta R \sim 0.3 R_{\rm J} \left( 1 + \frac{P_{\rm DD}}{L_{\rm eq}}\tau \right)^{\alpha} \ ,
    \label{inflate}
    \end{equation}
    where the factor 0.3 accounts for inflation from stellar radiation alone and $\alpha \sim 0.2$. This form is in good agreement with the more detailed numerics in Spiegel \& Burrows (2013) for arbitrary internal heat sources located deep inside the planet. The optical  depth $\tau \sim \kappa_{H} P/g$ (with $g=GM_{\rm P}/R_{\rm P}^2$) can reach values as high as $10^4$-$10^5$ for $T_{\rm eq.}\geq 2000$ K in a  regime where the H$^{-}$ opacity dominates. Accordingly, Fig.\ref{fig-anomaly} shows that our model provides conditions (i.e. the $P_{\rm DD}/L_{\rm eq}$ ratio) that allow variable amounts of inflation, upwards of 50\% and even as much as 100\%, depending on the optical depth. We have also verified that this estimate is in good agreement with Figure 6 in Miller et al. (2009), which is also based on an internal heating source. Our Fig. \ref{fig-anomaly} is similar to Fig. 4 in Ginzburg \& Sari (2015) which relies on more general arguments, supporting our contention that DD fusion can provide enough energy to inflate the planet as a true internal heat source.

 \subsection{Timescale of Radius Inflation}
      
 As mentioned in the introduction, the inflation should persist for times larger than  $\tau_{\rm KH}\sim 10^8$ yrs. To estimate the timescale for this effect,
we consider the total energy released by burning a mass $M_{\rm D}\sim 2.2\times 10^{-6} \eta_{\rm D, -4}\eta_{\rm ice, -1} M_{\rm c}$ of D in the core,
corresponding to the D density in Eq.~(\ref{metals}):

\begin{equation}
\label{eq:EDD}
E_{\rm DD} = Q_{\rm DD} \frac{M_{\rm D}}{m_{\rm D}} \sim 2.6\times 10^{42}\ {\rm erg}\ \eta_{\rm D, -4}\eta_{\rm ice, -1} \frac{M_{\rm P}}{M_{\rm J}}\ .
\end{equation}

 This energy will be gradually consumed at a non-constant rate determined by $P_{\rm DD}$. The timescale to reach the equilibrium radius $R_{\rm eq}$ (where $dR/dt\approx 0$) gives $P_{\rm DD}\approx L_{\rm eq}=4\pi R_{\rm eq}^2\sigma T_{\rm eff}^4$. Since we have $T_{\rm eff} \sim (200\,{\rm K})\,(M_{\rm P}/M_{\rm J})^{0.4}$, we estimate that 
  \begin{equation}
  \tau_{\rm DD}= \frac{E_{\rm DD}}{P_{\rm DD}}\sim 1.0\ {\rm Gyr}\ \eta_{\rm D,-4}  \eta_{\rm ice,-1} \left(\frac{M_{\rm J}}{M_{\rm P}}\right)^{0.6}\ ,
  \label{timescale}
  \end{equation}

where we have used $R_{\rm eq}$=1.2$R_J$. The above timescale of $\sim 1$ Gyr is an approximate lower bound for Jupiter-mass planets, given that the inflation effect can last longer as $T_{\rm eff}$ also decreases slowly during the evolution, though we assumed it constant. This estimate suggests that higher mass planets will have shorter inflation timescales, which agrees with Fig.\ref{fig-inflation}. The timescale also scales with $\eta_{\rm ice}$, which our evolution code also confirms - a larger ice content leads to a more persistent inflation effect. If ice occurs at the expense of metal content in the core, planets with lower metallicity (presumably derived from the parent star) would be more likely to be inflated as observed now, having a long-lived source for DD fusion. We remark on the observational trend with metallicity in \S \ref{sec:discussion}. We conclude that according to Eq.(\ref{timescale}) or Fig. \ref{fig-inflation}, Jupiter mass planets would inflate by about 20-50\%, with the effect clearly lasting for times of order $10^9$ yrs or more, which is larger than $t_{\rm KH}$. Besides the fusion power, inflation amount and timescale, there are other aspects of our model that we would like to explore.  In the next section, we discuss some interesting trends in the observational data that highlight possible sources of variability in our model. 
   
    \subsection{Dependence of Radius Anomaly on  $T_{\rm eq}$}
    \label{sec:trends}
    
     Based on observations of over 90 transiting giant exoplanets, Laughlin et al.  (2011) showed that the radius anomaly for $T_{\rm eq}> 1000$ K can be anywhere from 0 to 70\%, and that despite this variability, there appears to be a general rising trend between the radius anomaly $\Delta R$ = $R_{\rm obs}$-$R_{\rm pred}$  and effective planetary temperature $T_{\rm eq}$ with a best-fit dependence $\Delta R \propto T_{\rm eq}^{\alpha} \,{\rm with}\, \alpha=1.4\pm 0.6$. This cannot be explained by variations in planetary mass or stellar irradiation alone. Laughlin et al. (2011) used this correlation to argue for Ohmic dissipation from MHD effects in surface atmospheric flows as the preferred mechanism for inflation over kinetic heating. Since we are proposing quite a different mechanism (deep internal heating), it is interesting to see whether our model can accommodate this scaling. As mentioned in the introduction, subsequent work (Enoch et al. 2012; Bayliss et al. (2103); Weiss et al. (2013)) used a larger sample size and more accurate fits for subsets of the data that  take other planetary properties such as mass into account, leading to different exponents in the scaling relation. Therefore, we also wish to study the dependence of the radius inflation on planetary mass.
     
  Let us first focus on the $\Delta R$ scaling with the planetary equilibrium temperature $T_{\rm eq}$. The latter characterizes the incoming flux, and (roughly speaking) represents an average photospheric temperature given by $T_{\rm eq}=\left(R_{\ast}/(2a_{\ast})\right)^{1/2}T_{\ast}$ for zero albedo and orbital eccentricity, where $\ast$ denotes stellar numbers and $a$ is the semi-major axis (Laughlin et al. 2011). A general way to describe radius inflation from a deep energy deposition, assuming an $n=1$ polytrope for Jupiter-like planets, is given by Eq.~(\ref{inflate}), taken from Ginzburg \& Sari (2015), which agrees well with the numerical estimates presented in Spiegel \& Burrows (2013). Since the usual heat-blanketing effect from stellar irradiation is insufficient to explain the observed inflation, it follows that additional heat trapping must occur. This happens if $P_{\rm DD} \tau>L_{\rm eq}$ and is the effect described by Eq.~(\ref{inflate}). In our model, as shown by Fig. \ref{fig-anomaly}, we certainly have regimes where $P_{\rm DD}\tau/L_{\rm eq.} >>1$. Therefore, Eq.~(\ref{inflate}) becomes $\Delta R\propto (P_{\rm DD}\tau /L_{\rm eq.})^\alpha$.
 For a Rosseland opacity (i.e. a gray approximation) with $T_{\rm eq}\geq 2000$ K, the opacity $\kappa_{\rm H}$
     is dominated by that of  the negative Hydrogen ion H$^{-}$  and  scales as   $\kappa_{H}  \propto    T_{\rm eq}^s$
    with $s >>1$ and can be as high as $s= 9$ (e.g. Allard et al. 2001; see also Fig. 1 in Arras \& Bildtsten 2006). Since  $L_{\rm eq}\propto T_{\rm eq}^4$, this means
      that $\Delta R\propto  T_{\rm eq}^ {\alpha (s-4)}$. From Fig.~\ref{fig-pdd}, we see that $P_{\rm DD}$
    is more or less independent of $T_c$, the core temperature. Since $T_c$ and $T_{\rm eff}$ (the temperature at the base of the radiative layer) lie on the same adiabat (eg., Hubbard 1977; Guillot\& Showman 2002), we can find a relation between $T_c$ and $T_{\rm eq}$ assuming $T_{\rm eq}\approx 10T_{\rm eff}$ (eg., Spiegel \& Burrows 2013). For an $n=1$ polytrope, this yields $T_{\rm c}\sim 100 T_{\rm eff}$, which in turn implies $T_{\rm c}\sim 10 T_{\rm eq}$. This linear proportionality implies that $P_{\rm DD}$ does not scale significantly with $T_{\rm eq}$ and that, in our model, the scaling of radius inflation with $T_{\rm eq}$ is driven by the optical depth at which the heat is deposited, once the condition $P_{\rm DD}\tau/L_{\rm eq.} >>1$ is satisfied. Physically speaking, DD fusion provides a large enough power source, deep enough in the interior, to reinforce the slowing down of the planet's cooling due to trapping of stellar irradiation. Substituting typical values for $\alpha\sim 0.2$ (Ginzburg \& Sari (2015)) and taking into account the steep dependence of H$^-$ opacity on temperature with $s=9$, we get $\Delta R\propto T_{\rm eq}^{1.0}$, which agrees within systematics to the fit presented in Laughlin et al. (2011).  We stress that the application of Eq.(\ref{inflate}) to obtain the approximate scaling $\Delta R \propto T_{\rm eq}^{1.4\pm 0.6}$ claimed in Laughlin et al. (2011) requires $P_{\rm DD}\tau/L_{\rm eq.} >>1$, which in our model, holds true for more inflated planets when we apply the heuristic equations of Ginzburg \& Sari (2015). However, the fit in Laughlin et al. (2011) works better for moderately inflated planets (20\% or so). The difference could be explained by choosing a smaller value than $s$=9 for the $H^{-}$ opacity, which is on the higher end. A smaller opacity leads to less heat trapping and less inflation for the same $P_{\rm DD}$ and $\tau$, while giving a scaling that is weaker than $\Delta R \propto T_{\rm eq}^{1.0}$ but still consistent with the bounds in Laughlin et al. (2011) or Weiss et al. (2013). We emphasize that the fact that $P_{\rm DD}$ is at best weakly dependent on $T_{\rm eq}$ is important in getting the correlation index around 1.0, which further supports the choice of $U$ in the right range. Thus, we can be optimistic that the astrophysics implies a certain self-consistency in our model's assumptions.
    
 A related finding which is generally consistent with the scaling of  $\Delta R$ with $T_{\rm eq}$ is that planets receiving modest irradiation seem to lack inflation (e.g. Demory \& Seager 2011).  In our model, the condition for core erosion/solubility,  $T_{\rm c}   > T_{\rm ero}$ (with $T_{\rm ero}\sim 10^4$ K; Wilson \& Militzer (2012a\&b)), translates to a limit on the equilibrium temperature $T_{\rm eq} \geq 10^3\ {\rm K} \times (T_{\rm ero}/10^4 {\rm K})$, or in terms of irradiation $F_{\rm irr}$=$\sigma_{\rm SB}T_{\rm eq}^4 \geq  5.6\times 10^7$ erg cm$^{-2}$ s$^{-1} \times (T_{\rm ero}/10^4 {\rm K})^4$. This constraint on $T_{\rm eq}$, which is reflected in the observations, emerges as a necessary condition in our model but is not sufficient for inflation, since planets with higher metallicity (i.e. with reduced ice content) and/or smaller core will experience little to no inflation from DD fusion.
 
 \subsection{Dependence of Radius Anomaly on Planetary Mass}
    \label{sec:masstrends}
     
Now, let us look at the dependence on planetary mass in our model. Data from 119 transiting planets in different mass regions studied in Enoch et al (2012) shows that a different scaling of radius with $T_{\rm eq}$ for Saturn-mass planets, Jupiter-mass planets and high-mass planets yielded better fits to the radii of these planets. Bayliss et al. (2013) produced a density plot of 125 transiting exoplanets in short-period orbits that revealed an under-density of planets in the 1-2-1.6$M_{\rm J}$ range. Fig. 11 of Enoch et al. (2012) and Fig.1 of Laughlin et al. (2011) suggests higher mass planets are on average under-inflated. While more data is needed to clarify the trends, planetary mass clearly plays a role. Mass dependence enters into the DD fusion mechanism mainly through two factors. The first is the core density which depends on planetary mass. We use the fitted core EoS from Eq.(11) in Seager et al. (2007), with fit constants appropriate to an Fe-rich core (Table III of Seager et al. (2007)). Then,  pressure balance relates  core pressure to the stellar mass and radius which yields the required mass dependence of the core density for our evolution model. The second factor is the mean adiabat which connects the isothermal core  to $T_{\rm eff}$ which determines the cooling flux (ignoring the effect of stellar irradiation of the atmosphere). For a  mean adiabatic index of $\Gamma$=0.3 (eg. Hubbard et al. 1968) this yields  $T_{\rm eff}$ scaling as $M^{0.4}$. We also take into account the fact that sub-Jupiter mass planets are less compressible than super-Jupiters by choosing an appropriate polytropic $R$-$M$ relation, i.e, $n$=3/2 for high-mass planets and $n$=0.5-1.0 for lighter planets. Incorporating this dependence into our evolution model, Fig.\ref{fig-inflation} shows that sub-Jupiter (0.1-0.5 $M_{\rm J}$) and super-Jupiter mass planets (4-10 $M_{\rm J}$) are more inflated than planets in the mass range 1-2 $M_{\rm J}$. The reason for this trend within our model is that smaller mass planets have smaller $T_{\rm eff}$ which slows down the cooling while the larger mass planets have higher core mass and density leading to a larger $P_{\rm DD}$ that offsets the larger $T_{\rm eff}$. Comparing to Fig. 11 of Enoch et al. (2012) and Fig. 1 of Laughlin et al.(2011) suggests higher mass planets are on average under-inflated, which is not the case in our mechanism. A slight change in the mean adiabat dependence from the fully convective case of $n$=5/3 does lead to under-inflation of large mass planets in our model since then $T_{\rm eff}$ scales as $M^{0.5}$ or higher. In any case, it appears that our model needs to be re-examined for the high-mass region. We can conclude that our model certainly gives some of the right trends with mass, while a more comprehensive treatment of the interior structure of hot Jupiters needs to be taken up in future work to better confront our model with the observational trends.

 \section{Discussion \& Conclusions}
 \label{sec:discussion}
In the preceding sections, we have presented a model of DD burning in the core of extrasolar Jupiters in an attempt to explain the anomalously inflated radius of these planets.
In this section, we discuss some implications and limitations of the model. Since we have presented order of magnitude estimates and used a simple time-evolution model for the radius, we do not claim that DD fusion is $\it the$ mechanism for radius inflation. However, for hot Jupiters where other mechanisms may prove insufficient to inflate these planets, our estimates clearly show that DD fusion in the core could be an additional heat source that acts as a ``hot plate" inside the simmering planet, leading to over-inflation of their radius. For this work, we assumed that core erosion is induced by convection (e.g. Guillot et al. 2004). However, cracking and sputtering induced by DD fusion (e.g. Smentkowski 2000) may also lead to self-sustained core erosion (independent of convective plumes
 eroding the core). It may be that convection is needed early on to provide enough D-bearing material and then sputtering takes over. Once DD is triggered, such
 a self-regulating process can, in principle, take place between erosion (fuel supply) and burning (fuel consumption) leading to an equilibrium temperature in the $10^4$ K range, as required for efficient core erosion, according to studies by Wilson \& Militzer (2012a\&b).
 
 Core erosion is a necessary condition (to supply the projectiles) but this may not be sufficient if  not enough D in the core is available for burning. According to Eq.~(\ref{metals}), reduced ice content in the core (i.e. reduced D content) leads to lesser heat generation.  If the increase in core metallicity is at the expense of ice (and thus D content), then planets with metal-rich cores should experience less inflation despite ongoing core erosion. Guillot et al. (2006) did claim such a correlation, but suggested that planets around metal-rich stars are themselves metal-rich, making them denser and more compact (less inflation). They did not consider the possibility of internal heating from an icy core. In our model, metallicity can be further correlated with reduced ice content  (e.g. planets forming inside the snow line) which directly impacts the amount of inflation.  Combining the suggested explanation of Guillot et al. (2006) with ours, a natural connection emerges to the snow line in the disk: planets that formed beyond the snow line should carry healthy amount of ice in their cores making DD burning a more likely possibility when they migrate and park close to the star, and such hot Jupiters should show the most inflation. If the parent star metallicity and planet mass is known (eg., Miller \& Fortney (2011), one can then hope to use the radius trend with metallicity to constrain the planetÕs core metallicity, which is indirectly a parameter in our model through $\eta_{\rm ice}$. However, the issue of metallicity is complex, as even the correlation between parent star metallicity and planet metallicity has been questioned recently (Thorngren et al.(2015)), while other works Johnson et al. (2010) and Enoch et al. (2012) are more circumspect on the trend with metallicity than Guillot et al. (2006). Therefore, while our model would predict an anti-correlation trend, specially if metallicity is primarily residing in the core, we cannot infer any constraints on our model from the current trends in the data.
  
 Inflation seems to scale with the irradiation flux received by the planet (e.g. Laughlin et al.  2011), while in our model which does not take irradiation into account, the inflation arises from the high luminosity and energy deposition at large optical depth. Since irradiation is not sufficient to explain the observed inflation, the observational trend may also be suggesting that hotter surface temperature can induce conditions in the core favorable for erosion to be triggered. In that case, when there is sufficient D content in the core, ongoing DD burning would help sustain the planet's inflation and in some circumstances over-inflation. In extreme cases (e.g. $P_{\rm DD}/L_{\rm eq} > 0.01$ and $\tau> 10^5$),  inflation can lead to excessive mass-loss and possibly envelope stripping resulting in a naked ``super-Earth"  core
(the eroded core of the hot Jupiter).  

 \begin{figure}[t!]
\centering
\includegraphics[scale=0.9]{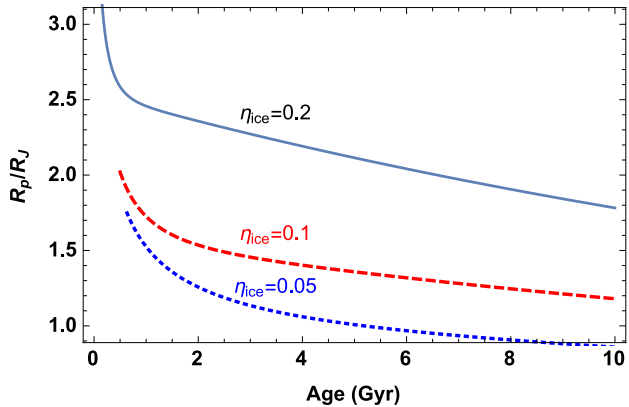}
\caption{Time evolution of $R_{\rm P}(t)$ including DD fusion in the planetary core with $\eta_{\rm ice}=\{0.2,0.1,0.05\}$ and no stellar irradiation
for a Jupiter-like planet  ($M_{\rm p}=M_{\rm J}$).~\\}
\label{fig-jupiter}
\end{figure}

The dependence on ice content in our model, implies that most inflated planets most likely formed beyond the ice line (i.e. accreted
the necessary ice along the rocky material) and migrated to park close to their parent star.  Once it collected enough  ice in its core, a young (hot) giant planet
is prone to DD burning as long as an erosion mechanism is at play. 
In principle, mechanisms other than trapping of stellar radiation (e.g. tidal heating) could be capable of inducing conditions in the core that trigger core erosion, leading to DD fusion. Thus inflated but colder Jupiters (at higher orbits) may exist, but be hard to find due to selection effects in transit detections. 
It is thus interesting to speculate on the implications of DD burning in our own Jupiter where the effects of irradiation can be ignored.  Fig. \ref{fig-jupiter} shows the evolution of the radius of a Jupiter-like planet as induced by DD fusion  for different ice content in the core.  The 
configuration with less ice content has  been slowly shrinking to Jupiter's current configuration purely due to cooling and eroding of the core. 
It suggests that  even for planets  far from their parent star, heat released from possibly ongoing DD fusion can keep the core temperature at values favorable for erosion to proceed and thus sustain the fusion. However, these planets will cool faster than the hot, parked, Jupiters and should deflate to  a radius close to the degenerate value after only a few billion years.  Another possibility, for example,  is that  planets which, long ago,  experienced DD fusion while parked  close to the parent star, could have migrated outward (e.g. due to planet-planet interactions),  and continue the DD burning process while losing the blanketing effect from irradiation. These planets may appear much less inflated than the parked ones but should nevertheless carry with them signatures of DD fusion such as an eroded core. In either case, in our model, this implies a reduced core in today's Jupiter which seems to be in agreement with recent models of the interiors of today's Jupiter  (e.g. Guillot et al. 2004;  Wilson \& Militzer 2012a\&b).

From the above points, and the arguments presented in the previous section, we conclude that screened DD fusion is a viable mechanism that can provide possible explanations for the two key facts listed in \S 1. We therefore suggest that screened DD burning operating in the deuterated core of giant planets is a plausible internal heat source and potential mechanism for the observed inflated radius. Based on estimates of the energy production in DD fusion at typical core temperatures, the amount of inflation and the trend with $T_{\rm eq}$ is broadly consistent with scaling relations extracted from observations. Scaling relations between radius expansion and mass are more subtle and requires further investigation in our model. Seismological models of Jupiter using inertial modes (e.g. Dintrans \& Ouyed 2001) can help probe the interior of the planet which hopefully would reveal signatures of a simmering DD burning in a reduced and slowly eroding core.   
  
\begin{acknowledgements}
R. Ouyed thanks T. Guillot and B. Hansen for discussions. 
The research  of R. O. is supported by an  operating grant from the
National Science and Engineering Research Council of Canada (NSERC). 
\end{acknowledgements}

\end{document}